\documentclass[osajnl,twocolumn,showpacs,superscriptaddress,11pt]{revtex4-1}
\usepackage{amsmath,amssymb,graphicx}
\usepackage[utf8]{inputenc}

\begin{document}


\title{Holographic microscopy reconstruction in both object and image half spaces with undistorted 3D grid}

\author{Nicolas Verrier}
\affiliation{Laboratoire Charles Coulomb  - UMR 5221 CNRS-UM2 Universit\'e Montpellier Place Eug\`ene Bataillon 34095 Montpellier, France}
\affiliation{Laboratoire Hubert Curien - UMR 5516-CNRS-Universit\'e Jean Monnet- 18 Rue du Professeur Beno\^it Lauras 42000 Saint-Etienne, France}
\author{Danier Alexandre}
\affiliation{Laboratoire Charles Coulomb  - UMR 5221 CNRS-UM2 Universit\'e Montpellier Place Eug\`ene Bataillon 34095 Montpellier, France}
\author{Gilles Tessier}
\affiliation{Holographic Microscopy Group, Neurophotonics Laboratory, CNRS UMR 8250, University Paris Descartes, Sorbonne Paris Cit\'e, 75006 Paris, France}
\affiliation{ESPCI ParisTech, PSL Research University, CNRS, Institut Langevin, 1, Rue Jussieu, F-75005 Paris, France}
\author{Michel Gross}\email{michel.gross@univ-montp2.fr}
\affiliation{Laboratoire Charles Coulomb  - UMR 5221 CNRS-UM2 Universit\'e Montpellier Place Eug\`ene Bataillon 34095 Montpellier, France}

%


\begin{abstract}
We propose an holographic microscopy reconstruction method, which propagates the hologram, in the object half space,  in the vicinity of the object. The calibration yields  reconstructions with an undistorted reconstruction grid i.e. with orthogonal $x$, $y$ and $z$ axis and constant pixels pitch. The method is validated with an USAF target  imaged by a $\times$60 microscope objective, whose holograms are recorded and reconstructed for different USAF locations  along the longitudinal axis: -75 to +75 $\mu$m.   Since the reconstruction numerical phase mask, the reference phase curvature and MO form an afocal device,    the reconstruction can be interpreted as occurring equivalently  in the object or in image half space.
\end{abstract}


\maketitle 

\section{Introduction}
In Digital Holography,  a camera records the interference
pattern of the object field wavefront with a known coherent
reference beam. This digital hologram is then used to
reconstruct numerically the  image of the object by back propagating
the measured  object field wavefront from the hologram to
the object \cite{schnars1994direct}. Many  methods have been proposed to reconstruct the holographic
image in free space  \cite{picart2008general}, like single Fourier
transform method \cite{schnars1994direct}, plane wave
expansion method with two Fourier transforms \cite{yun2004dms},
or adjustable magnification method \cite{zhang2004algorithm}.
Although digital
holographic microscopy is extensively used
\cite{Cuche99,ferraro2003dhm,charriere2006cri,garciasucerquia2006dlh,lee2007cat,sheng2006dhm},
very few papers describe the reconstruction procedure that must
be used in holographic microscopy.

Montfort et al. \cite{montfort2006npl}  proposed a
holographic microscopy reconstruction  algorithm, in which the field is propagated in free space from the camera to the image  of the object (conjugate of the object by the microscope objective MO) \cite{ferraro2003ciw,montfort2006npl,colomb2006npl,colomb2006tac,colomb2006apa}.
Various methods can be used then to compensate for the phase curvature of the lens, and for the tilt angle of  off-axis holography. Ferraro et al. \cite{ferraro2003ciw} use a reference hologram  as  phase mask.
Montfort et al., and Colomb et al. \cite{montfort2006npl,colomb2006npl,colomb2006tac,colomb2006apa} use a calculated phase mask that is defined by its  Zernike polynomial expansion, where coefficients are adjusted to optimize the contrast of the image.
Residual phase distortions can be then compensated by using a
 Zernike phase mask located in the plane of the object.
In most cases, these methods have been used
to measure the phase of ``flat objects'', e.g.  a biologic
sample between slide and cover slide or a micro lens device
\cite{zhang1998tdm,cuche1999sac,Cuche99,ferraro2003dhm,charriere2006cri,garciasucerquia2006dlh,lee2007cat}.
These methods  are well adapted to such cases, owing to these great efforts to compensate phase distortions and get a precise phase reference.
\begin{figure}
  \begin{center}
   \includegraphics[width=8.4cm]{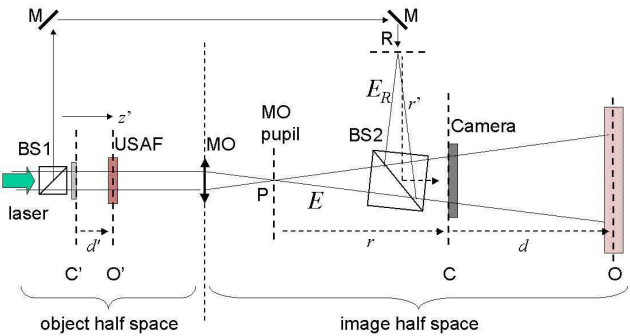}
    \caption{Typical holographic microscopy setup. USAF: USAF target located in plane O'; MO: microscope objective that images the USAF target in plane O; BS1, BS2: beam splitters; M: mirror. $E$ and $E_R$: signal and reference fields;  Note that BS2 is angularly tilted in order to perform off axis holography. } \label{fig_simple__setup}
  \end{center}
\end{figure}

To illustrate  reconstruction, let us consider a typical holographic microscopy setup (see Fig. \ref{fig_simple__setup}).  The object is located in plane O' and is imaged by the microscope objective  MO in plane O. The camera located in plane C records the hologram $H$ that is the interference pattern of the signal field $E$ with the reference field $E_R$. The purpose is to reconstruct the amplitude and the phase of the field scattered by the object in the object plane O',  and near this  plane (if the object is thick or if one want to measure the  wave field).
Before making  reconstruction, one need to calibrate the setup to get the reconstruction parameters, which depends on the  optical elements in the image half space: location of MO, BS2 and camera, BS2 angular tilt and direction and curvature of the reference beam (i.e. location of point R). Once the calibration of the setup is made, the reconstruction must be valid for any object. This means that one can  modify all the optical elements in the object half space (object and object location, direction and curvature of the illumination beam), but the optical elements in the image  half space must remains unchanged.

The  reconstruction methods mentioned above \cite{montfort2006npl,ferraro2003ciw,montfort2006npl,colomb2006npl,colomb2006tac,colomb2006apa}  made reconstruction in the image half space. This means that  the image of the object in plane O is reconstructed by propagating the hologram over a distance $d$ from C to O. This propagation is calculated in free space, by using either  the one Fourier Transform  (1-FFT) method \cite{schnars1994direct},  or the angular spectrum (2-FFT) method \cite{yun2004dms}. However, reconstruction in the  image half space becomes difficult when varying the location of the object along  $ z' $, or when imaging thick objects.
Indeed, the location of the image O and the reconstruction distance $ d = | \textrm {CO} | $ strongly depend on the object location $z'$, which complexifies the reconstruction process. If the object is located near plane C', which is imaged by  MO on  the camera (i.e. on plane C), we get $ d \simeq $ 0. The 2-FFT reconstruction method, which is well suited to small $ d $, should thus be preferred. Conversely, if the object is close to the MO focal plane, we get $ d \rightarrow \infty $, and the 1-FFT method should be used. This point tends to be problematic if we consider reconstruction of both amplitude and phase of of the light scattered by a thick 3D sample, or if one want to measure the wave field. As a matter of fact, it prevents experimental setup calibration to remain valid for the entire sample thickness. Making the setup calibration independent of the sample position according to MO is therefore an issue to tackle.

Here,  we propose to consider that the recorded signal  $|E+E_R|^2$  represents the hologram $H'$ of the field scattered by the object  in the plane of the image of the camera (i.e. in plane C'). The reconstruction is therefore made in the object half space, by propagating hologram $H'$   over a distance $d'=|\textrm{C'O'}|$ from plane C' to plane O'.
Since $d'$ is always small (a few microns), the 2-FFT method\cite{yun2004dms} is well adapted in all cases ($d\simeq 0$ or $d \simeq \infty$), which notably simplifies reconstruction algorithms.
Moreover, if the calibration is done well, this reconstruction can be made with orthogonal  $ x' $, $ y' $ and $ z' $  axes, while  keeping constant the pixel pitches in the $x'$ and $y'$ axes.

\section{Calculation of the hologram of the field in  plane C'}\label{section Calculation of the hologram of the field in  plane C'}

The hologram $H$ recorded by the camera in plane C is:
\begin{eqnarray} \label{Eq_H}
  H &=& |E+E_R|^2 \\
\nonumber &=& |E|^2+ |E_R|^2 + E E^*_R + E^*E_R
\end{eqnarray}
where $E$ and $E_R$ are the signal and reference fields in plane C. Because of the microscope objective MO, the curvature of the reference field and the off axis angular tilt, the phase of $H$ in plane C is not equal to the phase of the signal field in camera conjugate plane C', from which the object half space reconstruction will be made. Moreover, the +1 grating order term $ E E^*_R$ that is proportional to the field $E$ must be selectively filtered. To perform the reconstruction from the field $E$ in plane C', it is thus necessary to manipulate the data  to compensate for these phase effects and to select the $ E E^*_R$  term that is proportional to $E$ \cite{montfort2006purely}. This manipulation of the data must be made with  few calibration parameters that characterize the optical elements of the image half space part of the setup, and that do not  depends on the the object i.e. on the optical elements of the object half space.

\subsection{The phase curvature in the camera plane C }\label{section phase curvature in the camera plane C}

Let us  analyze first the phase curvature of $ E E^*_R$ in plane C. For that purpose, we will  consider a ``gedanken experiment" without object in which the illumination beam is  plane wave beam oriented along the  optical  axis $ z '$. The field $E$ is thus flat in plane C'.
%

This plane wave is focused by  MO at point P  in the MO pupil plane.
The  signal complex field $E$ exhibit thus a  phase factor equal to $e^{jk(x^2+y^2)/2r}$, where $k=2\pi/\lambda$ is the modulus of the wavevector in air, and where $r $ is the radius of curvature of the  wavefront that is equal to the MO pupil to camera distance $r=|\textrm{PC}| $.
Similarly, the  phase of the complex conjugate of the reference field  $ E_R^*$  is  $e^{-jk(x^2+y^2)/2r'}$ where $ r' = |\textrm{RC}| $ is the radius of curvature of $ E_R$ (with $ r' = \infty $ if the reference beam is a plane wave). This means that the phase of the +1 grating order term  $ E E^*_R$ in plane C is  $e^{+jk(x^2+y^2)/2r''}$ with   $1/r'' = 1/r - 1/r'$.

%
\begin{figure}
  \begin{center}
   \includegraphics[height=4cm]{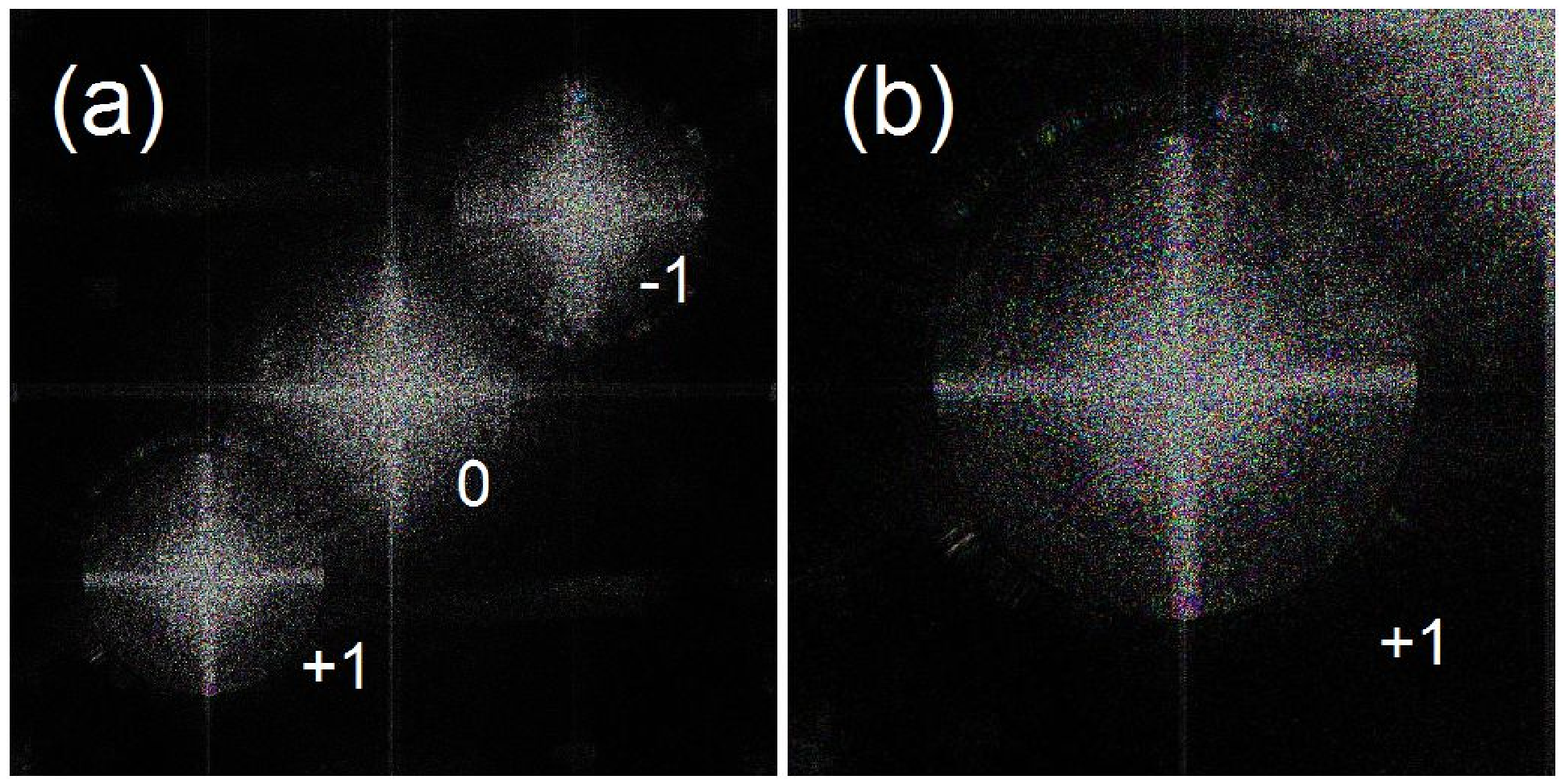}\\
      \includegraphics[height=4cm]{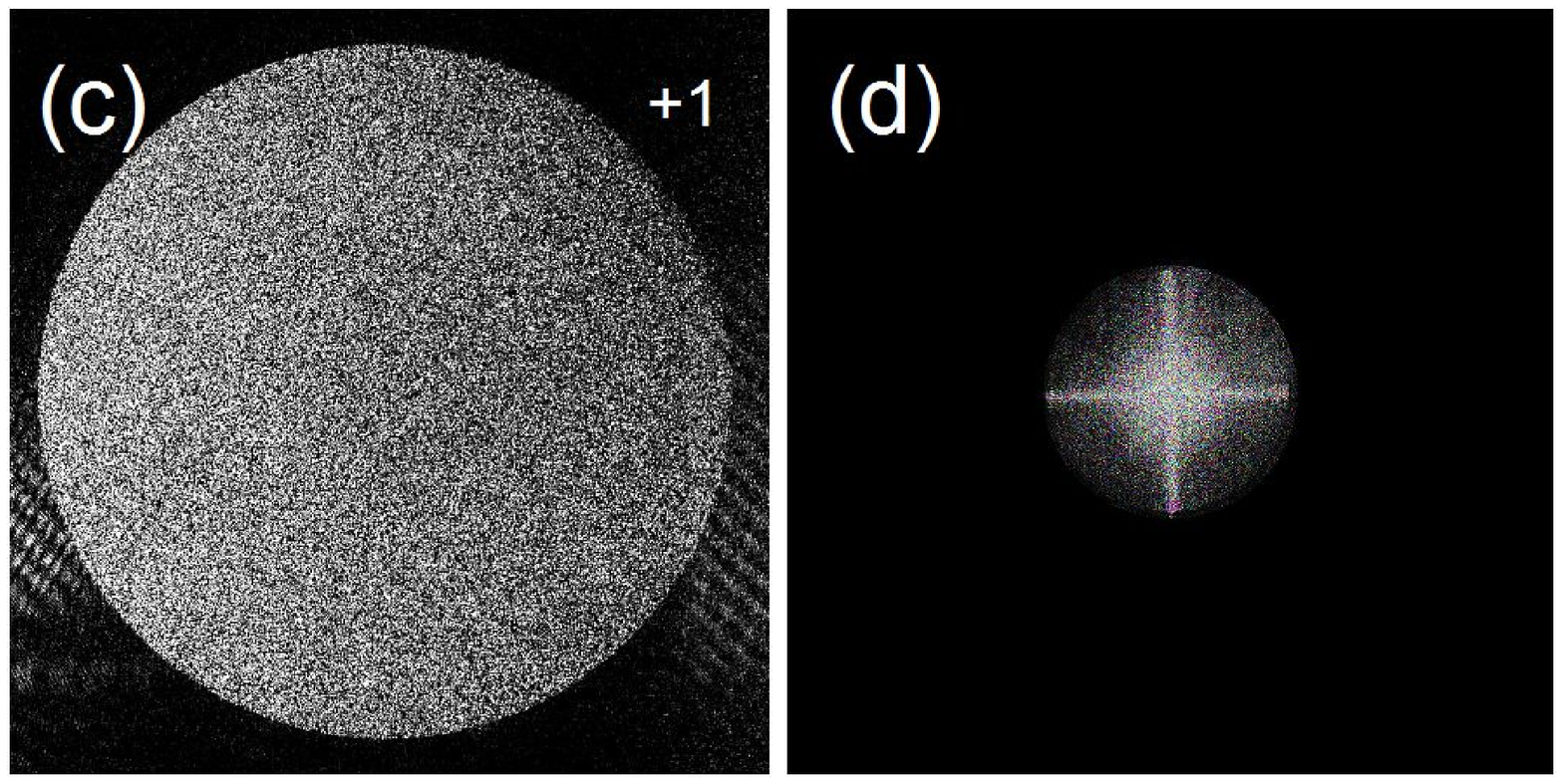}\\
     \includegraphics[height=4cm]{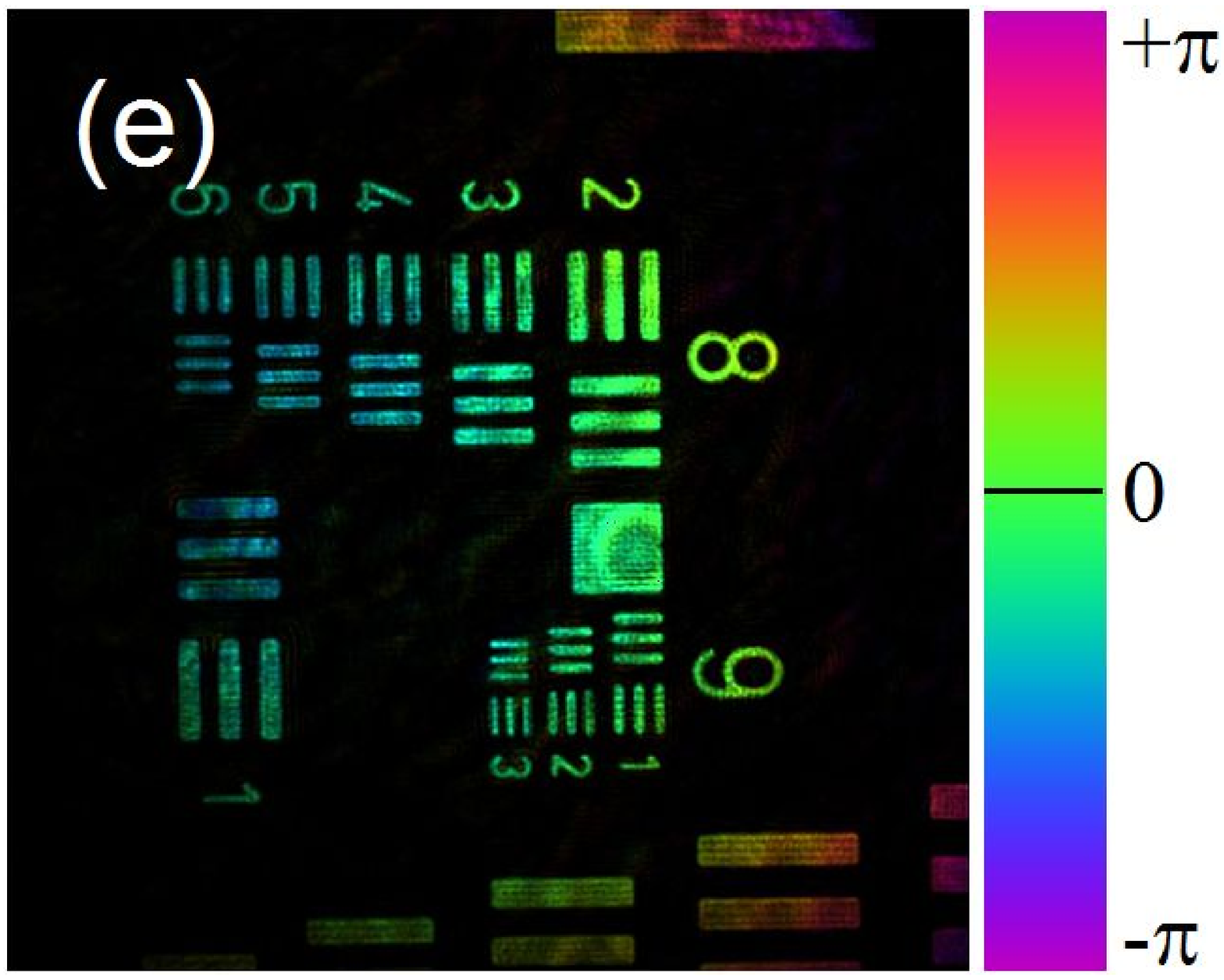}\\
 \caption{ (a)~Reconstructed USAF target hologram in the pupil
plane ${\tilde H_1}(k_x,k_y)$ ($1024\times 1024$ pixels). (b)~Zoom of the +1 grating order of
($512 \times 512$ pixels). (c) Pupil image obtained
with ground glass imaged with an $\times 10$ MO. (d)~Fourier space filtered
hologram ${\tilde H_2}(k_x,k_y)$ of the +1 grating order made with
a circular crop of the MO pupil image (radius $r_k$=162
pixels) followed by a translation of the selected zone in
the center of the Fourier space. (e)~Spatially filtered hologram
$H'_2(x, y)$ with proper phase correction. Arbitrary scale brightness is $|H_{...}|^2$, color is phase i.e.  $\arg H_{...}$.
} \label{fig_recons_29p}
  \end{center}
\end{figure}

\subsection{Phase curvature  correction  by reconstruction of the MO pupil}
%

Let us reconstruct the image $ {\tilde H}_1(k_x, k_y) $ of the MO pupil by using the  1-FFT  method  or Fourier reconstruction method\cite{schnars1994direct}. We have thus:
\begin{eqnarray}\label{EQ_schnars}
 {\tilde H_1}(k_x,k_y) = \textrm{FFT} \left[e^{-jk(x^2+y^2)/2r''} ~ H(x,y) \right]
\end{eqnarray}
where $e^{-jk(x^2+y^2)/2r''}$ is equal to $e^{-jk(x^2+y^2)/2r}$ $\times$  $e^{+jk(x^2+y^2)/2r'}$. The first phase factor $e^{-jk(x^2+y^2)/2r}$  is the 1-FFT   propagation kernel over a  distance $r$, while the second factor  $e^{+jk(x^2+y^2)/2r'}$ compensates   the curvature of $E_R^*$.  Since the back focal point P is located within the MO pupil plane,  the  propagation distance $r$ is equal to $|\textrm{PC}|$.
 This means  that the reconstruction phase factor $e^{-jk(x^2+y^2)/2r''}$ exactly compensates  the phase curvature $e^{+jk(x^2+y^2)/2r''}$ of $E E_R^*$ calculated in section \ref{section phase curvature in the camera plane C} .

To determine  $r''$, we propose to reconstruct  the MO pupil images $|{\tilde H_1}|^2$ by adjusting $r''$  so as to obtain the sharpest image of the pupil edge.
Since the we have validated our reconstruction with a USAF target located in near plane C',  we have reconstruct the pupil image  $|{\tilde H_1}(k_x,k_y)|^2$ obtained with the target in plane C', and displayed it on  Fig.\ref{fig_recons_29p} (a).  3 bright zones corresponding the +1, -1 and zero grating orders are visible. These zones are separated because of the off axis tilt angle. Here, the reconstruction parameter $r''$ has been properly adjusted, and the  pupil  edge of the +1 grating order zone is sharp as seen on the zoom displayed on Fig.\ref{fig_recons_29p} (b).

Nevertheless, the  pupil  edge is  dark on Fig.\ref{fig_recons_29p}~(a) and (b), because the  USAF target scatters light over quite narrow angles.  The adjustment of  $r''$ is thus quite difficult. It is easier to adjust  $r''$  with an  object that scatters more light, like a ground glass located in front of MO. Figure \ref{fig_recons_29p} (c) shows an example of a pupil reconstruction made with such a ground glass. Let us notice here  that  the adjustment of $r''$ can be made with the object of interest itself  without any previous calibration of the setup. This has been done in \cite{absil2010photothermal}.

\subsection{Spatial filtering and off axis phase correction }

To perform the spatial filtering that selects $E E_R^*$ and to  compensate  the off axis angular tilt, we propose to crop the data within the circular MO pupil image (which is sharp), and  to translate the cropped zone to the center of the calculation grid. The spatial filtering is made by the crop, while the off axis compensation is made by the translation.

The filtered and phase-corrected hologram of the pupil in the Fourier space  ${\tilde H}_2$ is thus:
\begin{eqnarray}\label{Eq_tildeH2}
 \nonumber {\tilde H}_2(\textbf{k}_{xy})  &= & {\tilde H_1}(\textbf{k}_{xy}+d\textbf{k}_{xy})~~~~~~~~  \textrm{if }~~ |\textbf{k}_{xy}| < r_k \\
                                &=& 0~~~~~~~~~~~~~~~~~~~~~~~~~~~  \textrm{if  not}
\end{eqnarray}
%
%
%
where $\textbf{k}_{xy}=(k_x,k_y)$, $d\textbf{k}_{xy}$ is the translation that moves the +1 pupil image in the center of the Fourier space, and $r_k$  the pupil radius. The corresponding real space hologram  ${ H}_2$ is:
\begin{eqnarray}\label{Eq_H2}
 H_2(x,y)  &= & \textrm{FFT}^{-1} \left[{ \tilde H}_2(k_x, k_y )~\right]
\end{eqnarray}
where $\textrm{FFT}^{-1}$ is the reverse 2D Fourier transform.

We have displayed $ |H_2(x,y)|^2$  on Fig. \ref{fig_recons_29p}(e) with  phase displayed  with color.   Since the USAF target is located in plane C', the image of the target is sharp. Moreover  the phase, which is displayed in color, is the same in all point of the image, like the phase of the field $E$ in plane C'.  This means that the phase has been properly corrected.

%
%

\subsection{Object/image half space holograms}

In the previous calculation, the holograms $H$, $\tilde H_1$, $\tilde H_2$  ... are  matrices of data, that represent the field either in the image or the object half space.

To avoid any confusion, we will consider that $H$, $\tilde H_1$, $\tilde H_2$ ...  are the holograms, and $x,y, k_x, k_y$ the coordinates   in  the image half space.   The  pitch   is   $\Delta x$ for $x$ and $y$, and $\Delta k= 2\pi/(N \Delta x)$ for  $k_x$ and $k_y$   (where   $\Delta x$ is the camera pixel size, and $N$  the size of the calculation grid) .

On the other hand,    $H'$, $\tilde H'_1$,  $\tilde H'_2$ ... are the holograms, and  $x',y', k'_x$ and $k'_y$ the coordinates  in the object half space.  The  pitch is  $\Delta' x =  \Delta' x/G$ for $x'$ and $y'$, and  $\Delta' k = G  \Delta' k$ for $k'_x$ and  $k'_y$ (where $G \gg 1$ is the MO transverse gain from plane C' to plane C).
We have thus:
\begin{eqnarray}\label{Eq_H_H'}
 H'_{...}(x',y')&=&H_{...}(x,y) \\
\nonumber \tilde{H}'_{...}( k'_x,  k'_y)&=&\tilde{H}_{...}(k_x,k_y)
 \end{eqnarray}
 with $x'=x/G$, $  y'=y/G $ , $k'_x=k_x G$ and $k'_y=k_y G$.

\section{Reconstruction of the field in any plane}

\subsection{Reconstruction in the object half space}

 We have calculated the phase-corrected hologram $H'_2$ in plane C' in section \ref{section Calculation of the hologram of the field in  plane C'} by  Eq.\ref{Eq_H2} and Eq.\ref{Eq_H_H'}.
The image $H'_3(x',y',z')$ of the object  in any $z'$ plane is reconstructed by using the 2-FFT reconstruction method (also known as the convolution method)  \cite{yun2004dms} from  $H'_2(x',y')$. Since ${\tilde H}'_2= \textrm{FFT}( H'_2) $, we have:
\begin{multline}\label{Eq_H_3}
    H'_3(x',y',z') = \\
    \textrm{FFT}^{-1} \left[ e^{j(k_x'^2 + k_y'^2)z'/2k_{m}} {\tilde H}'_2(k'_x,k'_y)  \right]
 \end{multline}
where $e^{j(k_x'^2 + k_y'^2)z'/2k_{m}}$ is the 2-FFT propagation kernel over distance  $z'$.  The origin $z'=0$ corresponds thus to plane C'. Note that since propagation from C' to O' takes place in a medium (air, water or oil) of refractive index $n_{\rm m}$, $k$ has been  replaced by $k_{\rm m}=n_{\rm m} k$ in the  propagation kernel.

\subsection{Reconstruction in the image half space}


Equation (\ref{Eq_H_H'}) makes the correspondance of the holograms $H_{...}$ and $H'_{...}$ in planes C and C'. It can be generalized for any plane by:
\begin{eqnarray}\label{Eq_H_H'_bis}
 H'_{...}(x',y',z')&=&H_{...}(x,y,z) \\
\nonumber \tilde{H}'_{...}( k'_x,  k'_y,z')&=&\tilde{H}_{...}(k_x,k_y,z)
 \end{eqnarray}
where   $z$ is a formal  coordinate that is not equal to $d=|\textrm{CO}|$.  We can now calculate  $H_3(x,y,z)$
%
%
by using a relation formally equivalent to  Eq. (\ref{Eq_H_3}). We get:
\begin{eqnarray}\label{Eq_H_3_image_half_space}
 \nonumber   H_3(x,y,z) = \textrm{FFT}^{-1} \left( e^{j(k_x^2 + k_y^2)z/2k} {\tilde H}_2(k_x,k_y)  \right)\\
 \end{eqnarray}
 %
Since  propagation in image half space is made in air, we have chosen in  Eq. (\ref{Eq_H_3_image_half_space}) a propagation kernel that  involves $k$ (instead of $k_{\rm m}$).  Equations  (\ref{Eq_H_3}) and  (\ref{Eq_H_3_image_half_space})  yield exactly the same calculations on the discrete data of  the calculation grid if we have  
\begin{eqnarray}\label{Eq_x'_z_z'}
     z &=& z' G^2/n_m
\end{eqnarray}
This equation defines the formal coordinate $z$.

\subsection{The afocal device that makes the  reconstructions in the object and image half space equivalent}
\begin{figure}
  \begin{center}
   \includegraphics[width=8.4cm]{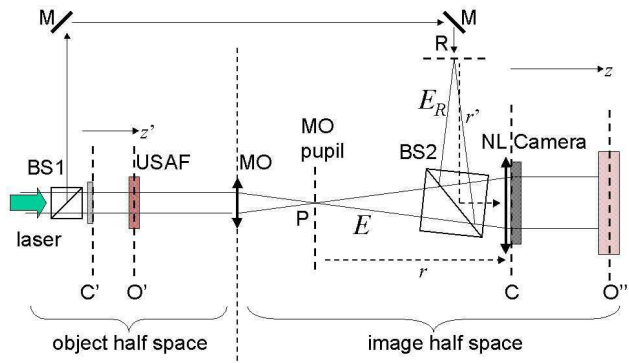}
    \caption{Reinterpretation of the holographic microscopy setup. USAF: USAF target located in plane O'; MO: microscope objective; NL: numerical lens of focal $r$ located in the camera plane C; MO + NL: afocal optical device  that images the USAF target in plane O''. } \label{fig_simple__setup1}
  \end{center}
\end{figure}

The  reconstruction made in the image half space by Eq. \ref{Eq_H_3_image_half_space} and Eq. \ref{Eq_x'_z_z'}  can be reinterpreted  quite simply.
The phase curvature of the reference beam (phase factor $e^{jk(x^2+y^2)/2r'}$) combined with the numerical phase factor $e^{jk(x^2+y^2)/2r''}$ used to image the MO pupil (and to correct the phase in plane C') is equivalent to a Numerical Lens (NL) of focal length $r$ (phase factor $e^{jk(x^2+y^2)/2r}$) located in the camera plane C (see Fig. \ref{fig_simple__setup1}). Since the MO and NL have a common focal point P, MO and NL form an  afocal optical device, with an overall transverse gain $G$ and a longitudinal gain ${G^2}/{n_{m}}$. Performing the reconstruction either in the object or  image half space is thus totally equivalent.

The reconstruction made in the image half space with the afocal optical device  is close to the reconstruction called  ''Image Plane Approach''  by Monfort et al. \cite{montfort2006purely}. Nevertheless, these authors consider a varying magnification factor, which involves explicitly the size of the  image of the object in plane O, and which depends  strongly on the location of the object. But the varying Monfort et al. magnification and the varying object  magnification  (plane O' to image  plane O: see Fig.\ref{fig_simple__setup}) compensate yielding a constant  magnification (object in plane O' to reconstructed image in plane O'': see  Fig.\ref{fig_simple__setup1}), which is simply  equal to G.
%
%

\section{Experimental validation}\label{section Experimental validation}

To validate our reconstruction method, we have made a test experiment with a $\lambda=$785 nm laser (Sanyo DL-7140-201) by imaging an USAF target with an oil immersion ($n_m=1.518$) Microscope Objective  MO (NA=1.4 $\times$60) and a holographic setup built by modifying a commercial  upright microscope. A detailed description of the setup is given in reference
 \cite{verrier2014laser}. The holograms are recorded
for $m = 0...59$ different positions $z'_m$ of the target along $z'$,  the positions being adjusted
manually with steps of $\Delta z'=2.5~\mu$m. The sharpest image
(nearly focused without any correction) corresponds to
location $m = 29$. The holograms are recorded with a  PCO Pixelfly camera
($1280\times 1024$ square pixels of size $\Delta x = 6.7 ~\mu$m), and the  measured data are  cropped into a $1024 \times 1024$ calculation grid.

\begin{figure}[h]
  \begin{center}
   \includegraphics[width=4.1cm]{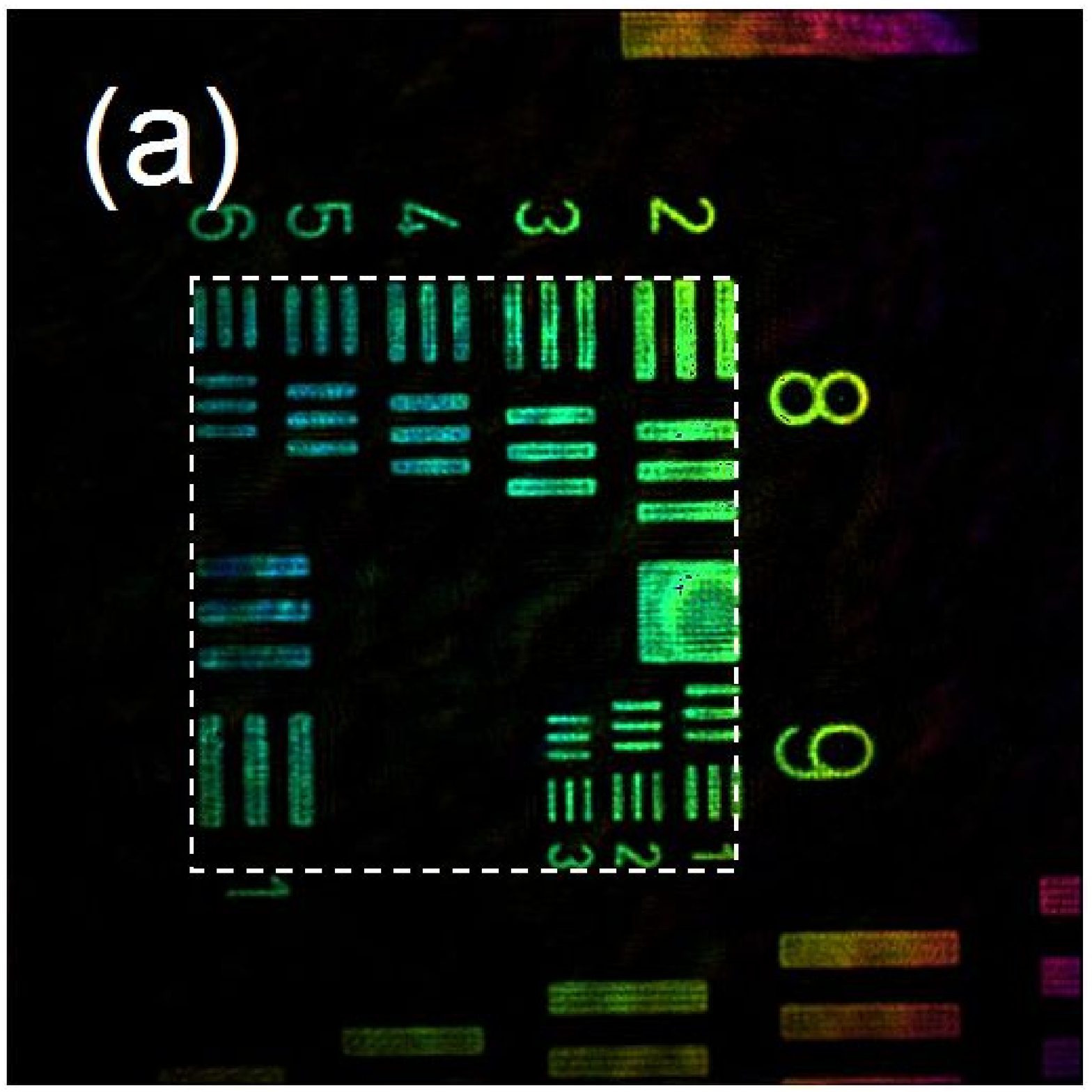}
      \includegraphics[width=4.1cm]{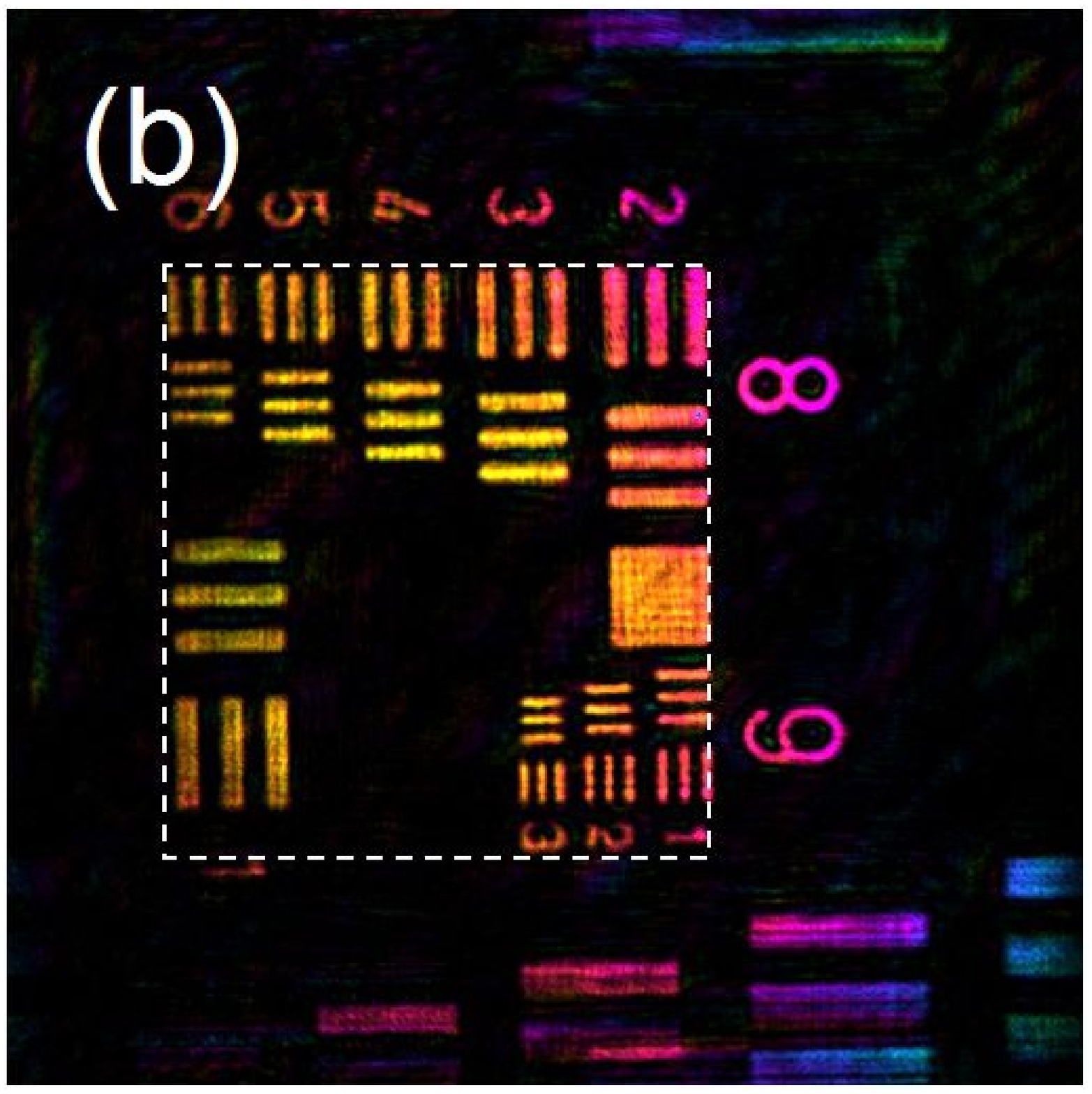}
 \caption{ (a) Reconstructed hologram $H'_3$ of the USAF target  for positions $m=29$ made with  $z'= 0$ (a), and $m=0$ with $z' \simeq -72.5 \mu$m (b). Arbitrary scale brightness is intensity i.e. $|H'_3|^2$, color is phase i.e.  $\arg H'_3$. The white dashed line rectangles ($517\times 567$ pixels)  seen are visual guides, whose sizes are  exactly the same size, and positions slightly shifted.
} \label{fig_recons_29_0}
  \end{center}
\end{figure}

The hologram  $H'_3$ of  the USAF target at position $m=29$  is displayed on  Fig. \ref{fig_recons_29_0} (a).  Since the USAF target is nearly focused without any correction, reconstruction is made with $z'=0$.  Moreover, since the phase correction has been made, the phase of the illumination beam is flat i.e. the color is approximately the same within the whole USAF target. We have then used the $m=29$ USAF image, which is sharp, to calibrate $G$. We got $G= 74.61$.

We have then reconstructed the USAF target for every position, from $m=0$ to $m=59$. Figure \ref{fig_recons_29_0} (b) shows for example the  USAF  image for position $m=0$.
As expected, the size and location of the USAF image remains the same, showing that the $x$ and $y$ scales are conserved. This can be verified by comparing   Fig. \ref{fig_recons_29_0} (a) and (b). For positions $m=0$ (corresponding to $z' \simeq -72.5~ \mu$m) and  $m=29$ ($z' \simeq 0$), the  size of the USAF target  is exactly preserved (see white dashed rectangle) while the  position is slightly shifted.

\begin{figure}
  \begin{center}
   \includegraphics[width=8.4cm]{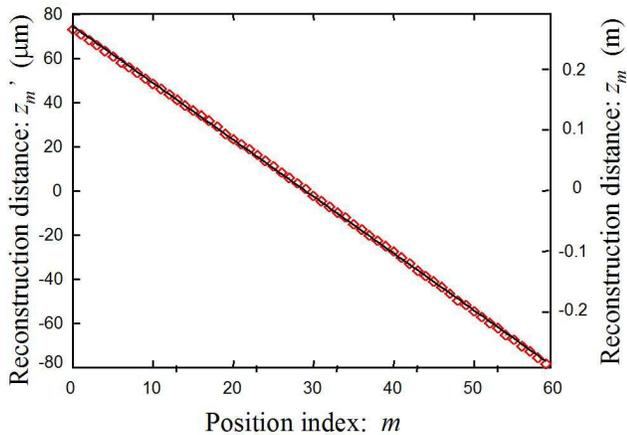}
    \caption{ Reconstruction distances $z'_m$ and $z_m$  as function of the USAF target location $m$ (red points) and best linear fit (black solid line). } \label{Fig_recons_curve}
  \end{center}
\end{figure}

For each position $m$, we have reconstructed the hologram
$H'_3(x', y', z'_m)$ by adjusting $z'_m$ (or $z_m$) so as to obtain the
sharpest image. As the USAF target is an ``amplitude
object'', we have adjust $z'_m$ by using the ''focus plane detection
criteria'' of Dubois et al. \cite{dubois2006focus}. Since the position and size of the image remain the same,  this automatic adjustment of $z'_m$ works well. Figure \ref{Fig_recons_curve} shows
the reconstruction distances $z'_m$ obtained by this method. As can be
seen, $z'_m$ varies linearly with  $m$ with
a slope $dz'_m/dm = -2.56 ~\mu$m which corresponds to the mechanical displacement, and with an offset $
m=28.97$ that is nearly equal to  the location $m=29$,  where the USAF target is on focus on the  camera.
%
%

\section{Remark on the calibration of the setup}

In section \ref{section phase curvature in the camera plane C}, we have calibrated the setup i.e. determine $r''$ and $d\textbf{k}_{xy}$ by reconstructing the image of the pupil  with an object that scatters enough light.  $r''$ is adjusted to get the sharpest pupil edges, and  $d\textbf{k}_{xy}$ by translating the pupil  in the center of the Fourier space.

In section \ref{section Experimental validation}, we have seen that the the size and location (in $x$ and  $y$ directions) of the reconstructed image of the USAF target do not change by moving the target  along $z'$. This property provides a second calibration method. One can move an object along the $z'$ direction and consider  2 different positions the object, for example $z'=-72.5 \mu$m, and $z' =0$. One can then record the holograms   and reconstruct the images of the object for these 2 positions,   as done in Fig.\ref{fig_recons_29_0} (a) and (b).
 One can then adjust  $r''$  in order to get the same  size for the 2 reconstructed images, and adjust  and $d\textbf{k}_{xy}$ to get the same positions.
In the example of Fig.\ref{fig_recons_29_0}, the adjustment of $r''$ is nearly perfect  (since the USAF sizes are exactly the same on (a) and (b) ), while the adjustment of $d\textbf{k}_{xy}$ could be improved (since the positions are slightly shifted).
Our experience is that second  calibration method (size and position of 2 reconstructed images with different $z'$)   yields more precise calibration than the first method (pupil sharpness and pupil translation).

The two proposed calibration methods    do not require a flat phase illumination  like the methods based on the phase flatness of the reconstructed image \cite{colomb2006apa}.
%
%
Note also that  the reconstruction direction $z'$ depends on the calibration method  which is used to get $d\textbf{k}_{xy}$.
If the phase flatness is used,  the $z'$ axis is the direction of the illumination beam.
If the translation of pupil image is used (first calibration method), $z'$ is parallel to the line that joins  the   pupil center to the camera center.
If  the position of  reconstructed  image of the object (USAF target) is used (second calibration method),  $z'$ is parallel with the  $z$ motion. This  last method is  the best method when a modified commercial microscope is used to make holography, since the $z'$ axis coincides with the microscope $z$ axis.

\section{Remark on construction with tube lens}

All the results presented here remains valid when a tube lens  placed between the pupil  P and the beam splitter BS2.    In the ''gedanken experiment'' with a plane wave illumination beam,  the tube lens changes the curvature of the field $E$ in plane C , i.e. $r$. The calibration parameter $r''$,  and the   NL focal lens $r$ are thus both changed. By using the second calibration method,  $r''$ is adjusted  so that the size of the reconstructed image  object is not dependent  on the object  position $z'$. This means that the  ensemble of  lenses  MO+tube lens+ NL still constitute an afocal device, with transverse and longitudinal magnifications  $G$ and  $G^2/n_m$.

\section{Conclusion}

We have proposed a holographic microscopy reconstruction method which propagates the hologram in the object half space, in the vicinity of the object.
Since the reconstruction phase, the reference phase curvature and MO form an afocal
device, the reconstruction can be interpreted as occurring equivalently in the object or in image half
space.

The reconstruction method has been  validated with a USAF target which has been  imaged with a high numerical aperture microscope objective MO for different locations of the target  along the longitudinal axis $z'$.
We have verified that the reconstruction is made with orthogonal axes $ x' $, $ y' $ and $ z' $, and that the pixel pitches   $x' $, $ y' $ and $ z' $ do not depend on the USAF target location. The  experimental test  has been  made with a plane illumination beam oriented along the $z$ axis, and we have verify that the reconstructed phase (colors on Fig. \ref{fig_recons_29_0} (a) and (b) ) varies slowly with $x,y$. The proposed reconstruction is  compatible with spherical or dark field illumination  ~\cite{atlan2008heterodyne,absil2010photothermal,warnasooriya2010imaging,verpillat2011dark}.

Two  calibration methods, which do not require plane wave illumination have been proposed. The first method is based on the reconstruction of the MO pupil, with an object that scatters  light.  This method can be used with holograms of interest which have recorded without proper calibration of the setup, and has been used in that case \cite{absil2010photothermal}. The second method, which is more precise, requires to translate an object along the $z'$ axis and to  record two  holograms for two different $z'$.

%
%
%
%
%

\section*{Acknowledgments}
We acknowledge funding by the Agence Nationale pour la Recherche, ANR Blanc, Simi 10, 3D BROM (ANR-11-BS10-0015).



\end{document}